\documentstyle[preprint,tighten,aps,floats,epsf,psfig,amssymb]{revtex}

\begin{document}

\def\spose#1{\hbox to 0pt{#1\hss}}
\def\ltapprox{\mathrel{\spose{\lower 3pt\hbox{$\mathchar"218$}}
 \raise 2.0pt\hbox{$\mathchar"13C$}}}
\def\gtapprox{\mathrel{\spose{\lower 3pt\hbox{$\mathchar"218$}}
 \raise 2.0pt\hbox{$\mathchar"13E$}}}

\draft

\title{ 
Dynamic structure factor of the Ising model with purely relaxational dynamics.
}
\author{
  \\
  {\small 
          Pasquale Calabrese,${}^{a,}$\cite{PC-email} 
          Victor Mart\'\i n-Mayor,${}^{b,}$\cite{VM-email} 
          Andrea Pelissetto,${}^{c,}$\cite{AP-email}  
          Ettore Vicari${}^{d,}$\cite{EV-email}  
}
\\[3mm]
  {\small\it ${}^a$ Scuola Normale Superiore and
              INFN, Sezione di Pisa,
              I-56100 Pisa, ITALIA} \\
  {\small\it ${}^b$ Departamento de F\'\i sica Teorica I, 
    Universidad Complutense de Madrid, } \\
  {\small\it   E-28040 Madrid, ESPA\~{N}A}  \\
  {\small\it ${}^c$ Dipartimento di Fisica, Universit\`a di Roma La Sapienza} 
         \\
  {\small\it and 
              INFN, Sezione di Roma,
              I-00185 Roma, ITALIA} \\
  {\small\it ${}^d$ Dipartimento di Fisica, Universit\`a di Pisa and 
              INFN, Sezione di Pisa,
              I-56100 Pisa, ITALIA} \\
}

\maketitle

\begin{abstract}
We compute the dynamic structure factor for the Ising model 
with a purely relaxational dynamics (model A). 
We perform a perturbative calculation in the $\epsilon$ expansion, 
at two loops in the high-temperature phase and at one loop in the 
temperature magnetic-field plane, and a Monte Carlo simulation in the 
high-temperature phase. We find that the dynamic structure factor 
is very well approximated by its mean-field Gaussian form up to 
moderately large values of the frequency $\omega$ and momentum $k$. 
In the region we can investigate, $k\xi \lesssim 5$, $\omega \tau \lesssim 10$,
where $\xi$ is the correlation length and $\tau$ the zero-momentum 
autocorrelation time, deviations are at most of a few percent. 
\end{abstract}


\newpage

\section{Introduction}

The dynamic structure factor $C(k,\omega)$
is a physically interesting quantity that can be directly
measured in scattering experiments.
Indeed, in neutron-scattering experiments
and in Born approximation, $C(k,\omega)$ is proportional 
to the cross section for inelastic scattering with momentum transfer
$k$ and energy transfer $\omega$. 
At a continuous phase transition the structure factor shows 
a universal scaling behavior
that depends on the dynamic universality class of the system.
In this paper we consider the dynamic universality class
of the three-dimensional Ising model with purely relaxational
dynamics without conservation laws, 
which is also known as model A, see Ref.~\cite{HH-77}.
This dynamic universality class should be appropriate to describe
the dynamic critical properties of uniaxial magnetic systems in which
the energy is not conserved due to the coupling to phonons and of 
alloys such as $\beta$-brass at the order-disorder transition;
see, e.g., Ref.~\cite{HH-77}. Note that this universality class does not 
describe the dynamic behavior of simple fluids and mixtures at the liquid-vapor
or mixing transitions because of additional conservation laws ~\cite{HH-77}.
The model-A dynamics for the Ising universality class may also be relevant 
for the dynamics of quarks and gluons at finite temperature and finite 
baryon-number chemical potential $\mu$. 
Indeed, using quantum chromodynamics, which is the current theory 
of strong interactions, one can argue that in the $T-\mu$ plane there exists
an Ising-like continuous transition 
at the endpoint of a first-order phase transition line \cite{HJSSV-98,BR-99}.
Model A (or model C if one takes into account the baryon-number
conservation \cite{BR-00}) should describe
the critical dynamics at this critical point (see also Ref.~\cite{BPSS-01}). 

In this paper we compute the structure factor 
$C(k,\omega)$ for the three-dimensional Ising
universality class with purely relaxational dynamics (model A) 
in equilibrium. 
We consider the $\epsilon$-expansion and compute $C(k,\omega)$ to two
loops in the high-temperature phase and to one loop in the whole 
temperature magnetic-field plane. In the high-temperature phase
we also perform a Monte Carlo simulation, 
using the standard Ising model and the Metropolis dynamics.
We find that, for moderately large $k$ and $\omega$, 
$C(k,\omega)$ is very well approximated by its mean-field (Gaussian) expression.
In the high-temperature phase the field-theoretical analysis and 
the simulation show that corrections to the mean-field behavior 
are less than 1\% for $k \xi \lesssim 5$ and $\omega \tau \lesssim 10$, 
where $\xi$ is the correlation length and 
$\tau$ is the zero-momentum autocorrelation time.
In the low-temperature phase, on the basis of a 
one-loop field-theoretical analysis, we expect 
slightly larger corrections, but still of the order of a few percent.
Note that our study concerns the scaling behavior of $C(k,\omega)$
in equilibrium, but it should be observed that similar conclusions have been 
obtained for the nonequilibrium dynamics in which one quenches a 
disordered system at $T_c$ \cite{CG-02}. 

The paper is organized as follows.
In Sec.~\ref{sec2} we define the quantities 
that are computed in the following sections. We report a list of definitions
together with some properties that are used in the calculation.
In Sec.~\ref{sec3} we present our field-theoretical results, obtained using 
the general formalism of Refs. \cite{MSR-73,BJW-76,DP-78}.
Sec.~\ref{sec4} is devoted to the presentation of the Monte Carlo results.
In the Appendix we report some technical details.

\section{Definitions and basic observables} \label{sec2}

In this paper we consider the equilibrium dynamics for 
an Ising-like theory with scalar order parameter
$\varphi(r,t)$ at temperature $T$ in the presence of a
time- and space-independent external (magnetic) field
$H$. We consider the connected two-point correlation function of the 
order parameter
\begin{equation}
G(r,t_1-t_2) \equiv  \langle \varphi(r,t_1) \varphi(0,t_2) \rangle_{\rm conn},
\end{equation}
where we have assumed to be in equilibrium, so that the correlation function
depends only on the difference $t_1 - t_2$.
Then, we define its Fourier transform $\widetilde{G}(k,t)$ with respect to $r$, 
\begin{equation} 
\widetilde{G}(k,t) = \int {d^dr}\,e^{ik\cdot r} G(r,t),
\end{equation}
and the structure factor $C(k,\omega)$ 
\begin{equation}
C(k,\omega) = \int dt\, e^{i\omega t} \widetilde{G}(k,t).
\label{Cdef}
\end{equation}
Here, we do not write explicitly the dependence on $T$ and $H$ that is 
always understood in the notation.
Near the critical point correlations develop both in space and time. 
They can be characterized in terms of the second-moment correlation length
$\xi$ and of the zero-momentum integrated autocorrelation time $\tau$ 
defined by
\begin{eqnarray}
\xi^2 &\equiv& {1\over 2 d \chi} \int d^d r\, |r|^2 G(r,0) = 
    - {1\over \chi} \left. {\partial \widetilde{G}(k,0) \over \partial k^2} 
      \right|_{k^2 = 0} , 
\label{xidef}
\\
\tau &\equiv& {1\over 2\chi} \int dt\, \widetilde{G}(0,t) = 
          {1\over 2\chi} C(0,0),
\label{taudef}
\end{eqnarray}
where $\chi \equiv \widetilde{G}(0,0)$ is the static magnetic susceptibility.
As is well known, for $T\to T_c$ ($T_c$ is the critical temperature) 
and $H\to 0$, $\xi$ and $\tau$ diverge.
In the absence of a magnetic field, 
\begin{equation}
\xi \sim |T-T_c|^{-\nu}, \qquad \qquad \tau \sim |T-T_c|^{-z\nu} \sim \xi^z,
\end{equation}
where $\nu$ is the usual static exponent and $z$ a dynamic exponent that 
depends on the considered dynamics. 
The static exponents for the three-dimensional 
Ising universality class are very well
determined \cite{GZ-98,CPRV-99,BST-99,Hasenbusch-01,BC-02,CPRV-02}, 
see Ref.~\cite{review} for a review. Present-day lattice studies 
give estimates that can be summarized as follows \cite{review}: 
$\gamma = 1.2372(5)$, $\nu = 0.6301(4)$, $\eta = 0.0364(5)$, 
$\alpha = 0.110(1)$. The exponent $z$ depends on the dynamics. 
For model-A dynamics, estimates of $z$ in three dimensions 
have been obtained by employing
several methods. There exist field-theoretical perturbative calculations 
in different schemes \cite{BDJZ-81,AV-84,PIF-97} and Monte Carlo analyses
that determine $z$ by studying the equilibrium dynamics 
at $T_c$ in finite volume \cite{WL-91,Heuer-92}, 
damage  spreading \cite{Gro-95,Grassberger-95}, the critical relaxation 
from an ordered state \cite{SK-96,Stauffer-97}, hysteresis scaling \cite{ZZ-98},
and the short-time critical dynamics \cite{JMSZ-99}.
For experimental determinations see, e.g., 
Refs.~\cite{Belanger-etal-88,Hutchings-etal-72}. 
The exponent $z$ turns out to be  slightly larger than two.
For instance, $z\approx 2.017$ from the fixed-dimension field-theoretical
expansion \cite{PIF-97}, $z\approx2.02$ from an analysis
interpolating the $4-\epsilon$ and $1+\epsilon$ expansions
\cite{BDJZ-81}, and $z\approx 2.04$ from Monte Carlo simulations.

Near the critical point correlation functions show a scaling behavior. 
For the static structure factor, neglecting scaling corrections,
we have \cite{FB-67,TF-75}
\begin{equation}
{\chi \over \widetilde{G}(k,0)} \approx g_{\rm stat}(y;x),
\label{gstat}
\end{equation}
where $x\equiv a_0 (T-T_c) M^{-1/\beta}$, $y\equiv k^2 \xi^2$, 
$M \equiv \langle \varphi\rangle$ is the time-independent 
(we only consider the equilibrium dynamics) static magnetization, and 
$a_0$ is a normalization factor that is fixed by requiring that 
$x=-1$ corresponds to the coexistence line. The magnetization 
$M$ is related to $T$ and $H$ by the equation of state, which, 
in the critical limit, can be written in the scaling form
\begin{equation}
H = b_0 M^\delta f(x),
\end{equation}
where $b_0$ is a nonuniversal constant which is fixed by the condition 
$f(0) = 1$. 

The function $g_{\rm stat}(y;x)$ has been extensively studied, 
both in the high-temperature \cite{CPRV-98,MMPV-02} 
and in the low-temperature phase \cite{CDK-75}; see Ref. \cite{review}
for an extensive review. 
In the high-temperature phase, the static function 
$g_{\rm stat}^+(y)$ is known to $O(\epsilon^3)$ \cite{Bray-76},
and satisfies $g_{\rm stat}^+(y) = 1+y+O(\epsilon^2 y^2)$. Its
small-momentum expansion in three dimensions has been accurately determined 
using high-temperature expansion techniques, see, 
e.g., Refs.~\cite{CPRV-02,review}, finding 
\begin{equation}
g_{\rm stat}^+(y) = 
1 + y - 0.000390(6) \,y^2 + 0.0000088(1) \,y^3 + O(y^4).
\end{equation} 
There are also precise estimates 
of the equation-of-state scaling function $f(x)$ 
\cite{GZ-97,CPRV-99,CPRV-02,EFS-02}.

Eq.~(\ref{gstat}) can be extended to finite values of $t$. In the critical
limit we can write
\begin{equation}
{\chi\over \widetilde{G}(k,t)} \approx g(y,s;x)
\end{equation}
with $s\equiv t/\tau$. We can also define a scaling function for the 
structure factor:
\begin{equation} 
{C(k,\omega)\over 2\tau \chi} \approx {\cal C} (y,w;x),
\end{equation}
where $w\equiv \omega \tau$ and 
\begin{equation}
{\cal C}(y,w;x) = {1\over2} \int ds\, e^{-iws} [ g(y,s;x)]^{-1} .
\end{equation}
We also define an integrated autocorrelation time at momentum $k$,
\begin{equation}
\tau(k) \equiv {1\over2} 
 \int dt\, {\widetilde{G}(k,t)\over \widetilde{G}(k,0)} = 
  {C(k,0)\over 2\widetilde{G}(k,0)},
\end{equation}
and an exponential autocorrelation time
\begin{equation}
\tau_{\rm exp}(k) \equiv - \lim_{|t|\to\infty}
      {|t|\over \ln \widetilde{G}(k,t)},
\end{equation}
which controls the large-$t$ behavior of $\widetilde{G}(k,t)$: 
$\widetilde{G}(k,t) \sim \exp[-|t|/\tau_{\rm exp}(k)]$ for $|t|\to \infty$. 
In the scaling limit, neglecting scaling corrections,
\begin{eqnarray}
{\tau(k)\over \tau} &\approx& {\cal T}(y;x) = 
{\cal C}(y,0;x) g_{\rm stat} (y), \\
{\tau_{\rm exp}(k)\over \tau(k)} &\approx& {\cal T}_{\rm exp}(y;x) = 
{1\over |w_0(y;x)| {\cal C}(y,0;x) g_{\rm stat} (y)},
\label{Texp-def}
\end{eqnarray}
where $\pm i w_0(y;x)$ are the zeros of $[{\cal C}(y,w;x)]^{-1}$ 
at fixed $y$ and $x$ on the imaginary $w$-axis that are nearest to the 
origin $w=0$. 

For a Gaussian free theory we have
\begin{equation}
C(k,\omega)|_{\rm Gaussian} = 
  {2 \Omega \chi m^2\over \Omega^2 (m^2 + k^2)^2 +\omega^2},
\label{CGauss}
\end{equation}
where $\Omega$ is an Onsager transport coefficient and 
$m \equiv 1/\xi$. It follows
\begin{eqnarray}
&& [{\cal C}(y,w;x)]^{-1} = (1 + y)^2 + w^2, \nonumber \\
&& {} [{\cal T}(y;x)]^{-1} = 1+y, \nonumber \\
&& {\cal T}_{\rm exp}(y;x) = 1.
\label{Gaussian-beh}
\end{eqnarray}
For $y\to 0$ and $w\to 0$ the above-defined scaling functions have a 
regular behavior and one can write 
\begin{eqnarray}
{} [{\cal C}(y,w;x)]^{-1} &=& (1 + y)^2 + w^2 + 
       \sum_{m,n=0} c_{n,m}(x) y^n w^{2m}, \nonumber \\
{} [{\cal T}(y;x)]^{-1} &=& 1+y +\sum_{n=0} t_{n}(x) y^n, \nonumber \\
{\cal T}_{\rm exp}(y;x) &=& 1 + \sum_{n=0} t_{{\rm exp},n}(x) y^n, 
\label{smallyw-exp}
\end{eqnarray}
with $c_{0,0}(x) = 0$ because of the definition of $\tau$. 
The expansion coefficients $c_{n,m}(x)$, $t_{n}(x)$, and 
$t_{{\rm exp},n}(x)$ parametrize the deviations from the Gaussian 
behavior (\ref{Gaussian-beh}) in the low-frequency and low-momentum regime.

At the critical point, $T=T_c$, $H=0$, the structure factor obeys the 
scaling law
\begin{equation} 
C(k,\omega) = {1\over \omega^{(2 -\eta + z)/z}} f_C(\omega k^{-z}),
\end{equation}
with $f_C(\infty)$ finite, which implies that, for $y\to\infty$, $w\to\infty$
keeping $u\equiv w y^{-z/2} = u_0 \omega k^{-z}$ fixed, we have 
\begin{equation} 
{\cal C}(y,w;x) = {f_0\over w^{(2 -\eta + z)/z}} f_C(u/u_0),
\end{equation}
where $f_0$ is a normalization constant.

For large $w$ at $y$ and $x$ fixed we have 
\begin{equation}
{\cal C}(y,w;x) \approx c_\infty(y;x) w^{-(2-\eta+z)/z},
\label{CalC-largew}
\end{equation}
where $c_\infty(y;x)$ is finite and $x$-independent for $y\to\infty$.
The large-frequency behavior of the structure factor allows us
to compute the nonanalytic small-$s$ behavior of 
$g(y,s;x)$ at $y$ and $x$ fixed. We obtain for $s\to 0^+$ \cite{intom}
\begin{equation}
[g(y,s;x)]^{-1}_{\rm nonanalytic} = 
  g_0(y;x) s^{(2-\eta)/z},
\label{g-smallt}
\end{equation}
where 
\begin{equation}
g_0(y;x) = - {1\over \pi} c_\infty(y;x) \sin (\pi\rho/2) \Gamma(1-\rho), 
\end{equation} 
with $\rho \equiv 1 + (2-\eta)/z$.
Notice that, since $(2-\eta)/z\approx 0.96$, the 
nonanalytic small-$t$ behavior of 
$\widetilde{G}(k,t)$ turns out to be practically
indistinguishable from the analytic background.

\section{Field-theoretical results}
\label{sec3}

\subsection{Field-theoretical approach}

In order to determine the critical behavior of a purely relaxational 
dynamics without conservation laws, the so-called model-A dynamics, 
one may start from the 
the stochastic Langevin equation \cite{HH-77}
\begin{equation}
\label{lang}
\frac{\partial \varphi (r,t)}{\partial t}=-\Omega 
\frac{\delta \cal{H}(\varphi)}{\delta \varphi(r,t)}+\rho(r,t),
\end{equation}
where $\varphi(r,t)$ is the order parameter,   
$\cal{H}(\varphi)$ is the Landau-Ginzburg-Wilson Hamiltonian
\begin{equation}
{\cal H}(\varphi) = \int d^d x \left[
\frac{1}{2} (\partial \varphi )^2 + 
\frac{1}{2} r \varphi^2
+\frac{1}{4!} u \varphi^4  
- H \varphi \right],
\end{equation}
$\Omega$ a transport coefficient (cf. Eq. (\ref{CGauss})), and $\rho(t)$ a
Gaussian white noise with correlations
\begin{equation}
\langle \rho(r,t)\rangle = 0, \qquad\qquad
\langle \rho(r_1,t_1) \rho(r_2,t_2)\rangle = 
    \Omega \delta(r_1-r_2) \delta(t_1-t_2). 
\end{equation}
The  correlation functions generated by the Langevin equation (\ref{lang}) and
averaged over the noise $\rho$, 
can be obtained starting from the 
field-theoretical action \cite{MSR-73,BJW-76,DP-78} 
\begin{equation}
S(\varphi,\hat{\varphi})= \int d t d^d x\; \left[ 
\hat{\varphi} \frac{\partial \varphi}{\partial t}+
\Omega \hat{\varphi} {\delta \mathcal{H}(\varphi)\over \delta \varphi}
-\Omega \hat{\varphi}^{\,2}
-{\rm ln} J(\varphi) \right].
\label{MRS}
\end{equation}
The last term in the action is  an appropriate Jacobian term
that  compensates the contributions of self-loops 
of response propagators  \cite{BJW-76,DP-78}.
 
In order to perform the field-theoretical calculation it is useful to introduce
the response function $Y(r,t)$---it gives the linear response to an external
magnetic field---defined by 
\begin{equation}
Y(r,t_1-t_2) = \langle \hat{\varphi}(r,t_1) \varphi(0,t_2)\rangle,
\end{equation}
(again we have assumed to be in equilibrium so that time-translation invariance
holds), its Fourier transform $\widetilde{Y}(k,t)$ with respect to $r$, 
and its double Fourier transform $R(k,\omega)$ with respect to $r$ and $t$, 
defined as $C(k,\omega)$ in Eq.~(\ref{Cdef}). 
The response function and the two-point correlation function are 
strictly related. First, the zero-frequency response functions are related
to the static correlation functions,
\begin{equation}
\widetilde{G}(k,0) = \Omega R(k,0).
\label{GR-rel}
\end{equation}
Moreover, because of the fluctuation-dissipation theorem that holds for the 
equilibrium dynamics, we have 
\begin{equation}
\label{fdth}
\omega C( k,\omega)=2 \Omega \, {\rm Im}\, R( k,\omega).
\end{equation}
Also the response function $R(k,\omega)$ shows a scaling behavior and 
one can write 
\begin{equation}
{\chi\over \Omega R(k,\omega)} \approx r(y,w;x),
\end{equation}
neglecting scaling corrections. The function $r(y,w;x)$ is such that 
\begin{eqnarray}
&& r(y,0;x) = 1 + y + O(y^2), \nonumber \\
&& r(0,w;x) = 1 - i w + O(w^2), \nonumber \\ 
&& {} [r(y,-w;x)]^* = r(y,w;x).
\label{normal-r}
\end{eqnarray}
Then, it is easy to show by using Eqs.~(\ref{GR-rel}) and (\ref{fdth}) that
\begin{eqnarray}
r(y,0;x) &=& g_{\rm stat}(y;x), \nonumber \\
{\cal C}(y,w;x) &=& - {{\rm Im}\, r(y,w;x)\over 
       w |r(y,w;x)|^2}.
\label{relr-C}
\end{eqnarray}
For a Gaussian theory
\begin{equation}
r(y,w;x) = 1 + y - i w.
\label{r_Gauss}
\end{equation}
The behavior of $r(y,w;x)$ for small $w$ and large $w$ is similar to that 
of ${\cal C}(y,w;x)$. For small frequencies and momenta, 
the scaling function has a regular expansion in powers of $w$ and $y$:
\begin{equation}
r(y,w;x) =
g_{\rm stat}(y;x) - i w
\left[ 1 +  
\sum_{n,m} r_{n,m}(x) (iw)^m y^n \right] ,
\label{r-smallyw}
\end{equation}
where the coefficients ${r}_{n,m}(x)$ are real and parametrize the 
$w$-dependent deviations from the Gaussian behavior (\ref{r_Gauss}). 
For $w\to\infty$ at fixed $y$ we have 
\begin{equation}
r(y,w;x) \approx r_\infty^+(y;x) (-i w)^{(2-\eta)/z}.
\label{r-largew}
\end{equation}

\subsection{Correlation functions in the disordered phase} \label{sec3.B}

\begin{table}
\begin{center}
\caption{\label{tabcnj} 
Numerical values of the coefficients $\bar{r}_{n,m}^+$ for $0\le m,n \le 3$. }
\begin{tabular}{c|cccc}
 $n \backslash m $  & 0 & 1 & 2 & 3 \\
\hline
0 & 0 & 1.03122$\times 10^{-3}$ &   6.19416$\times 10^{-5}$ &   6.51316$\times 10^{-6}$  \\
1 & $-$1.04876$\times 10^{-3}$  & $-$8.72163$\times 10^{-5}$ &
$-$1.15844$\times 10^{-5}$ &  $-$1.92466$\times 10^{-6}$\\
2 & 4.23375$\times 10^{-5}$   & 7.93211$\times 10^{-6}$ & 1.68104$\times 10^{-6}$  & 3.86236$\times 10^{-7}$\\
3 & $-$2.48539$\times 10^{-6}$ &   $-$7.51416$\times  10^{-7}$ &
  $-$2.2168$\times 10^{-7}$ &    $-$6.54471$\times 10^{-8}$\\
\end{tabular} 
\end{center}
\end{table}

In this Section we consider the equilibrium dynamics in the high-temperature
phase $H=0$, $T>T_c$. 
In order to determine the two-point correlation function,
we have computed the scaling function 
$r^+(y,w)$ (here and in the following we will not indicate $x$ and add instead 
a superscript ``+" to remind the reader that we refer to the high-temperature
phase) and we have then used Eq.~(\ref{relr-C}). 

A two-loop calculation in the framework of the $\epsilon$ expansion gives
\begin{equation}
r^+(y,w)
= g_{\rm stat}^+ (y) - i w  \left[ 1 + \epsilon^2 A(y,w) + O(\epsilon^3)
\right], 
\label{ayw}
\end{equation}
where $A(y,w)$ is reported in App.~\ref{appa}. Note that $A(0,0)=0$ 
and $A(y,-w)^* = A(y,w)$, as expected from Eq.~(\ref{normal-r}).
The static function 
$g_{\rm stat}^+(y)$ is known to $O(\epsilon^3)$ \cite{Bray-76},
and at order $\epsilon^2$ it reads
\begin{equation}
g_{\rm stat}^+(y) =1+y + 10^{-3}\,  \epsilon^2[-3.76012 y^2
+ 0.095966 y^3 -0.00407101 y^4+ O(y^5)]+O(\epsilon^3)\,.
\end{equation}
Expanding $A(y,w)$ in powers of $y$ and $w$ one obtains the coefficients
$r^+_{n,m}$ defined in Eq.~(\ref{r-smallyw}). We have 
$r^+_{n,m} = \epsilon^2 \bar{r}_{n,m}^+$, where 
$\bar{r}_{n,m}^+$ are reported in Table~\ref{tabcnj} for $n,m\le 3$.
The coefficients $\bar{r}_{n,m}^+$ 
are rather small, the largest ones being of order
$10^{-3}$, and decrease quite rapidly. The analysis of the 
coefficients of the expansion of $A(k^2,\omega)$ in powers of $k^2$ 
(at fixed $\omega$) shows the presence of a 
singularity  for $w = - 3 i$. Therefore, we expect asymptotically
\begin{equation}
\bar{r}^+_{n,m} \approx  {1\over3} \bar{r}^+_{n,m-1}.
\label{rel-r-w}
\end{equation}
We have verified numerically this relation,
although quantitative agreement is observed only for 
quite large values of $m$: for $n=0$, this relation is satisfied at the 
10\% level only for $m\ge 41$. Analogously, the
coefficients of the expansion of $A(k^2,\omega)$ in powers of $\omega$
become singular for $k^2 = -9$, so that asymptotically
\begin{equation}
\bar{r}^+_{n,m} \approx  - {1\over9} \bar{r}^+_{n-1,m}.
\label{rel-r-y}
\end{equation}
The behaviors (\ref{rel-r-w}) and (\ref{rel-r-y}) can be interpreted in terms
of the analytic structure of $R^+(k,\omega)$. If one considers the structure 
factor, it is well known \cite{FS-75,Bray-76} 
that the nearest singularity \cite{foot-3pc} appearing 
in $[\widetilde{G}(k,0)]^{-1}$ is the three-particle cut 
at $k=\pm 3im_{\rm exp}$, where 
$m_{\rm exp}$ is the mass gap of the theory. Since in the critical limit
$m_{\rm exp} \xi \approx 1$ \cite{review} with very small corrections
(more precisely $m_{\rm exp} \xi - 1 = - 2.00(3)\cdot 10^{-4}$,
see Ref. \cite{CPRV-02}),
the nearest singularity to the origin appearing in $g^+_{\rm stat}(y)$ 
corresponds to $y\approx -9$. In view of relation (\ref{rel-r-y}), 
it is natural
to conjecture that the same behavior holds for $R^+(k,\omega)$, so that 
Eq.~(\ref{rel-r-y}) should approximately hold for the 
three-dimensional coefficients 
${r}^+_{n,m}$ and not only for their two-loop approximation. 

Relation (\ref{rel-r-w}) is 
consistent with the idea that the three-particle cut also controls
the small-$w$ behavior. In this case it is natural to conjecture 
that the coefficients of the expansion of $[R^+(k,\omega)]^{-1}$ in powers of 
$k^2$ have a singularity for 
$\omega = - 3 i/\tau_{\rm exp}(0)$. Thus, turning to the scaling function 
$r^+(y;w)$, we expect a singularity at 
$w = - 3 i \tau/\tau_{\rm exp}(0) \approx -3 i$,
since, as we shall see, in the critical limit 
$\tau/\tau_{\rm exp}(0) \approx 1$. 
Therefore, we expect relation (\ref{rel-r-w}) to be a general 
property of the three-dimensional coefficients ${r}^+_{n,m}$.

This discussion indicates that ${\cal C}(y,0)$ and its 
$w$-derivatives at $w = 0$ have a convergent 
expansion in $y$ for $|y|\lesssim 9$ and analogously that 
${\cal C}(0,w)$ and its $y$-derivatives at $y=0$ have a convergent
expansion for $|w|\lesssim 3$. Mathematically, this  does not tell us 
much about the convergence of the double expansion which requires to 
know the singularity structure for both 
$y,w\not = 0$. At two loops, one can easily verify from the exact expression
that ${\cal C}(y,w)$ has a convergent double expansion 
in the whole region $|w| < 3$, $|y| < 9$, and it is sensible to conjecture that 
the same is true for the exact expansion.
From the results of Table~\ref{tabcnj}, one sees 
quite clearly that the 
response function $R^+(k,\omega)$ is well described by the Gaussian 
approximation for $|w| \lesssim 3$ and $|y| \lesssim 9$. Deviations should be 
smaller than 1\% in this region. This result is very similar to that obtained 
for the static structure factor: in that case high-temperature expansions
and Monte Carlo simulations \cite{MMPV-02} 
show that the deviations from the 
Gaussian behavior are less that 0.3\% for $y \lesssim 9$.

We now consider the large-frequency behavior. 
At order $\epsilon^2$ the function $r_\infty^+(y)$
defined in Eq.~(\ref{r-largew}) turns out to be constant and given by 
\begin{equation}
r_\infty^+(y) = 1  + 0.00538992  \epsilon^2 +  O(\epsilon^3).
\end{equation}
Again the correction term is quite small. 

\begin{table}
\begin{center}
\caption{\label{tabpnj} 
Numerical values of the coefficients $\bar{c}_{n,m}^+$ for 
$0\le n,m\le 3$.} 
\begin{tabular}{c|cccc}
 $n \backslash m $  & 0 & 1 & 2 & 3 \\
\hline
0 & 0&0.00212438& $-$0.000075868 & 1.22037 $\times 10^{-6}$\\
1 & 0.00104876&0.000951544&9.73059$\times10^{-7}$&$-$1.87657$\times{10}^{-7}$\\
2 & $-$0.0170494&$-$0.000075777&1.13334$\times{10}^{-6}$&
       $-$1.13131$\times{10}^{-8}$\\
3 & 0.00106254&3.43201$\times{10}^{-6}$&$-$2.22548$\times{10}^{-7}$&
      1.03457$\times{10}^{-8}$\\ 
\end{tabular}
\end{center}
\end{table} 
Using the fluctuation-dissipation theorem (\ref{relr-C}),
we obtain for the scaling function ${\cal C}^+(y,w)$:
\begin{equation}
[{\cal C}^+(y,w)]^{-1} = g_{\rm stat}(y)^2 + w^2  + 
\epsilon^2 E(y,w) + O(\epsilon^3),
\end{equation}
where 
\begin{equation}
E(y,w) = 2 w (1+y) {\rm Im}\, A(y,w) + \left[ w^2 -
(1+y)^2\right] {\rm Re}\, A(y,w).
\end{equation}  
We can then obtain the small-$w$ and small-$y$ behavior. For the 
coefficients $c_{n,m}^+$, see Eq. (\ref{smallyw-exp}), we obtain
$c_{n,m}^+ = \epsilon^2 \bar{c}_{n,m}^+$, where 
the constants $\bar{c}_{n,m}^+$ are 
reported in Table \ref{tabpnj} for $n,m\le 3$. 

Again, we should note that the coefficients $\bar{c}_{n,m}^+$ are very small 
and show the same pattern observed for $\bar{r}_{n,m}^+$. We expect 
that ${\cal C}^+(y,w)$ has singularities at $y=-9$ and $w=\pm 3 i$, so that 
$|c_{n,m}/c_{n+1,m}| \approx |c_{n,m}/c_{n,m+1}| \approx 9$. Thus, 
in complete analogy with what observed for the static structure factor
and $R^+(k,\omega)$, 
the dynamic ${C}^+(k,\omega)$ is essentially Gaussian in the region 
$y\ltapprox 9$ and $|w|\ltapprox 3$. 

We also compute the large-frequency behavior. 
For the coefficients $c_\infty^+(y)$ and $g_0^+(y)$, see 
Eqs.~(\ref{CalC-largew}) and (\ref{g-smallt}), we obtain 
\begin{eqnarray}
c_\infty^+(y) &=&  1 - 0.00538992 \epsilon^2 + O(\epsilon^3),  \\
2 g_0^+(y) &=&  1+0.00136716 \epsilon^2 + O(\epsilon^3), 
\label{flb}
\end{eqnarray}
where, since the corrections are very small, one may simply set $\epsilon=1$ 
to obtain a three-dimensional numerical estimate. 
Therefore, for large $w$ we predict
${\cal C}^+(y,w) \approx 0.995/w^{1.95}$, which is not very different from the
purely Gaussian behavior ${\cal C}^+_{\rm Gauss}(y,w) \approx 1/w^2$. 
Thus, the
Gaussian approximation should  be a reasonably good approximation 
even outside the small-$w$
region, $w\ltapprox 3$, discussed above. Trusting the above estimate 
of $c_\infty^+(y)$ we find that ${\cal C}^+(y,w)/{\cal C}^+_{\rm Gauss}(y,w)
=1.12$, 1.25, 1.41 respectively for $w=10$, 100, 1000. Thus, quite 
large values of $w$ are needed in order to observe a significant difference.

Finally we compute the scaling function ${\cal T}_{\rm exp}^+ (y)$
defined in Eq. (\ref{Texp-def}).
For this purpose we need to compute $\tau_{\rm exp}(k)$ and therefore the 
large-$t$ behavior of $\widetilde{G}(k,t)$. Because of the 
fluctuation-dissipation theorem, it is equivalent to consider 
$\widetilde{Y}(k,t)$. For $y<3$ we obtain
\begin{equation}
\widetilde{Y}(k,t) \approx  e^{-s(1+y)}\{1-\epsilon^2 s A[y,-i(1+y)]\},
\end{equation}
where $s \equiv t/\tau$, while for $y > 3$ we have
\begin{equation}
\widetilde{Y}(k,t) \approx e^{-s(1+y)}+\epsilon^2
\frac{27}{8} {e^{-s(y/3 + 3)} \over s^4 (y-3)^2 (y+9)^3} 
   [1 + O(s^{-1/2})] + O(\epsilon^3)
\end{equation}
For $y<3$ the correction term exponentiates as expected, and as a consequence 
we obtain
\begin{equation}
{\cal C}^+ (y,0) [g_{\rm stat}^+(y)]^2 {\cal T}_{\rm exp}^+ (y)=
1 + \epsilon^2 A[y,-i(1 + y)] + O(\epsilon^3).
\label{Texp-ymin3}
\end{equation}
On the other hand, for $y > 3$ the correction term decreases with 
a different exponential factor which dominates for large values of $t$,
suggesting that, at leading order in $\epsilon$,  
$\tau_{\rm exp}(k)/\tau = (y+9)/3$.
In other words, the interaction turns on a new singularity (a three-particle 
cut) that becomes the leading one for $y$ large enough. However, this is 
not the end of the story. Indeed, by considering graphs in which one 
recursively replaces each line with a two-loop watermelon graph one 
obtains contributions to $\widetilde{Y}(k,t)$ decreasing as 
$\exp[-s (3^{-n} y + 3^n)]$ ($3^n$-particle cut) which would be more 
important for $y$ large enough. These singularities will not probably be the 
only ones, since we also expect a 5-particle cut, a 7-particle cut, etc.
On the basis of these results, we expect ${\cal T}_{\rm exp}^+ (y)$ to have 
several singularities on the positive real $y$ axis and to become eventually
infinite as $y\to \infty$. This is not unexpected since, for $y\to\infty$,
$R^+(k,\omega)$ behaves as $\omega^{-(2-\eta)/z}$ and therefore has a branch
cut starting at $\omega=0$.

For $y<3$, we can use Eq.~(\ref{Texp-ymin3}) to compute the 
coefficients $t_{{\rm exp},n}^+$ defined in Eq.~(\ref{smallyw-exp}). 
We obtain, at order $\epsilon^2$,
$t_{\rm exp,0}^+ = 0.00110075 \epsilon^2$,
$t_{\rm exp,1}^+ = 0.00337789 \epsilon^2$, 
$t_{\rm exp,2}^+ = 0.000217173 \epsilon^2$, etc.
The coefficients decrease as $t_{\rm exp,n}^+/t_{\rm exp,n+1}^+\approx 3$,
which reflects the presence of a singularity at $y=3$.
Again, for $y<3$ the deviations from a purely Gaussian behavior are very small.

\subsection{Correlation function in the $(t,H)$ plane} \label{sec3.C}

\begin{table}
\begin{center}
\caption{\label{tabLT} 
Numerical values of the coefficients $\bar{r}_{n,m}^-$, $\bar{c}_{n,m}^-$,
$\bar{t}_n^-$, and $\bar{t}_{{\rm exp},n}^-$  
for $n\le 2$ and $m\le 3$. }
\begin{tabular}{c|cccc}
 $m$  & 0 & 1 & 2 & 3 \\
\hline
$\bar{r}_{0,m}^-$ &   0   & ${1\over 48}$ & ${1\over 192}$ & ${1\over 640}$ \\
$\bar{r}_{1,m}^-$ &  $-{5\over192}$ & $-{3\over 320}$ &
               $-{7\over 1920}$ &   $-{1\over 672}$  \\
$\bar{r}_{2,m}^-$ & ${11\over 1920}$ & ${29\over 8960}$ & 
          ${37\over 21504}$ &   ${115\over 129024}$ \\
\hline
$\bar{c}_{0,m}^-$ &   0   & ${3\over 64}$  &  $-{17\over 1920}$  & 
             $\frac{69}{71680}$ \\
$\bar{c}_{1,m}^-$ &  ${5\over 192}$ & ${7\over 1920}$ & 
               $\frac{221}{71680}$ &  $-\frac{251}{322560}$  \\
$\bar{c}_{2,m}^-$ & $\frac{19}{640}$ &   $-\frac{743}{107520}$  & 
$-\frac{1}{4032}$& $\frac{949}{2838528}$ \\
\hline
$\bar{t}_m^-$ & 0 & ${5\over 192}$ & ${23\over 1920}$ & $-{697\over215040}$ \\
\hline
$\bar{t}_{{\rm exp},m}^-$ &
${3\over 8} - {1\over 2} \ln 2$ &
$-{65\over 64}+{3\over 2} \ln 2$ &
${1551\over 640}-{7\over 2} \ln 2$ & $-\frac{422211}{71680}+\frac{17}{2}\ln 2$
\end{tabular}
\end{center}
\end{table}

In the presence of an external magnetic field $H$, 
a one-loop calculation gives
\begin{eqnarray}
\label{byw}
r(y,w;x) = g_{\rm stat}(y;x) - i w
\left[ 1 + \epsilon {2\over 3+x} B(y,w)
+O(\epsilon^2) \right],
\label{sec3C.eq1}
\end{eqnarray}
where $B(y,w)$ is defined in App.~\ref{appa} and 
\begin{eqnarray}
g_{\rm stat}(y;x)= 1 + y +  {2\epsilon \over 3+x} \left[
-1 - {y\over 12} + {\sqrt{4+y}\over 2 \sqrt{y}}\ln
    {\sqrt{4+y} + \sqrt{y}\over \sqrt{4+y} - \sqrt{y}}
\right]
+O(\epsilon^2).
\label{sec3C.eq2}
\end{eqnarray}
Note that the $O(\epsilon)$
correction vanishes for $x\to \infty$ in agreement with the results of the 
previous Section. Moreover, the $x$-dependence is very simple 
and in Eqs. (\ref{sec3C.eq1}) and (\ref{sec3C.eq2}) is always given
by the prefactor $2/(3 + x)$ that becomes 1 on the coexistence curve
$x=-1$. As a consequence, such a prefactor will always appear in this Section,
multiplying the low-temperature results that will be specified by adding 
a superscript ``$-$" to all definitions. Of course, such a simple $x$-dependence
does not hold at higher loops, as it can be seen for instance from the 
two-loop results of Ref. \cite{CDK-75} for the static structure factor.

One can easily derive the small-momentum and small-frequency behavior 
by expanding the function $B(y,w)$. The coefficients $r_{n,m}(x)$, 
see Eq.~(\ref{r-smallyw}), are given by 
\begin{equation}
r_{n,m}(x) = 2 \epsilon{ \bar{r}^-_{n,m}\over 3 + x} + O(\epsilon^2),
\end{equation}
where $\bar{r}^-_{n,m}$
are given in Table \ref{tabLT} for $m\le 3$ and $n\le 2$.

Again, we note that the corrections to the Gaussian behavior 
are small, although a factor-of-ten larger than the corresponding 
high-temperature ones. For instance, $\bar{r}^-_{0,1} \approx 0.02$ 
to be compared with $\bar{r}^+_{0,1} \approx 0.002$. Moreover, the coefficients
decrease slower with $n$ and $m$. This fact can be understood in terms 
of the singularities of the function $B(y,w)$. A simple analysis
shows the presence of singularities for $w = - 2 i$ and $y = - 4$, so that 
asymptotically
\begin{equation}
\bar{r}^-_{n,m} \approx {1\over2} \bar{r}^-_{n,m-1}, \qquad\qquad
\bar{r}^-_{n,m} \approx - {1\over4} \bar{r}^-_{n-1,m}. 
\label{rel-coeffr-LT}
\end{equation}
This behavior can be understood on general grounds. Considering the 
static structure factor, it is known that the nearest singularity in 
the low-temperature phase is the two-particle cut $k=\pm 2im_{\rm exp}$, 
so that $g^-_{\rm stat}(y)$ has a singularity for $y = - 4 (m_{\rm exp} \xi)^2
\approx - 4$, where we have used the fact that in the 
critical limit $m_{\rm exp} \xi \approx 1$ 
(more precisely, $m_{\rm exp} \xi \approx 0.96(1)$ \cite{FZ-98,CPRV-99}). 
As we did for the high-temperature phase we can thus conjecture that also 
the singularities of the dynamic functions are controlled by the 
two-particle cut. Therefore, we expect singularities for 
$y = - 4 (m_{\rm exp} \xi)^2 \approx -4$ and 
$w = - 2 i \tau/\tau_{\rm exp}(0) \approx -2i$, where we have used the 
fact that $\tau/\tau_{\rm exp}(0) \approx 1$, with corrections of order 
a few percent as discussed below, in the critical limit.
Therefore, Eq.~(\ref{rel-coeffr-LT}) should also approximately apply to the 
three-dimensional coefficients $r^-_{n,m}$. 

The above-reported discussion shows that in the region $|y|\lesssim 4$, 
$|w| \lesssim 2$ the response function can be reasonably approximated by
a Gaussian form. Note however that, while in the high-temperature phase 
corrections are expected to be less than 1\%, here deviations should be 
larger. 

We have also studied the large-frequency behavior.
The coefficient $r_\infty(y;x)$ turns out to be $y$-independent at one loop:
\begin{equation}
r_\infty(y;x)=1-{\epsilon\over 3 + x}+O(\epsilon^2). 
\end{equation}
Note that the correction is quite large and thus significant deviations
for the Gaussian behavior should be observed as soon as $w$ is large.

Using the fluctuation-dissipation theorem, we can compute at one loop
the scaling function ${\cal C}(y,w;x)$.
For the small-$w$, small-$y$ coefficients, we obtain 
\begin{eqnarray}
&& c_{n,m}(x) = {2 \epsilon\over 3 + x} \bar{c}^-_{m,n} + O(\epsilon^2), 
\nonumber \\
&& t_{n}(x) = {2 \epsilon\over 3 + x} \bar{t}^-_{n} + O(\epsilon^2).
\end{eqnarray}
The coefficients $\bar{c}^-_{m,n}$ and $\bar{t}^-_n$
are reported in Table \ref{tabLT} for $n\le 2$ and $m\le 3$. 

We have also investigated the large-frequency behavior. 
It is very simple to show, using the above-reported formulas, that at this 
order $c_\infty(y;x)=1/r_\infty(y;x)$ and  $g_0(y;x)=c_\infty(y;x)/2$.

Finally, we consider ${\cal T}_{\rm exp}(y;x)$. For this purpose we need to
compute the large-$t$ behavior of $\widetilde{Y}(k,t)$. We observe 
a behavior analogous to that observed in the high-temperature phase.
For $y<2$, 
\begin{equation}
{\cal C} (y,0;x) [g_{\rm stat}(y;x)]^2 {\cal T}_{\rm exp} (y;x)=
1 + {2\epsilon\over 3 + x} B[y,-i(1 + y)] + O(\epsilon^2),
\label{Texp-ymin2}
\end{equation}
while for $y> 2$ the two-particle cut contribution dominates so that 
$\tau_{\rm exp}(k)/\tau = 2 + y/2$. The discussion reported in Sec.~\ref{sec3.B}
can be repeated also here. One can easily identify diagrams that decrease 
as $\exp[-s(2^{-n}y+2^n)]$, indicating that ${\cal T}_{\rm exp} (y;x)$
has an infinite number of singularities on the $y$ axis and that it
diverges for $y\to \infty$.
For small $y$, we can use Eq.~(\ref{Texp-ymin2}) to compute the 
small-$y$ expansion coefficients $t_{{\rm exp},n}(x)$. We have
\begin{equation}
t_{{\rm exp},n}(x) = {2 \epsilon\over 3 + x} \bar{t}^-_{{\rm exp},n} + 
O(\epsilon^2).
\end{equation}
Numerical values are reported in Table \ref{tabLT}.
Note that $\tau_{\rm exp}(0)/\tau = 1 + t_{{\rm exp},0}^- \approx
1 + 0.0284\epsilon$, 
and thus we expect this ratio to be 1 with corrections of order of a few 
percent.

\section{Monte Carlo results} \label{sec4}

\begin{table}
\caption{Results of the Monte Carlo simulations.}
\label{tableMC}
\begin{tabular}{lcccc}
& (a) & (b) & (c) & (d) \\
$L$ & 64 & 64 & 128 & 128 \\
$\beta$ & 0.215 & 0.219 & 0.2204 & 0.2210 \\
$N_{\rm it}$ & 30$\times$10$^6$ & 8$\times$10$^6$ & 9$\times$10$^6$ & 
   4$\times$10$^6$ \\
$\xi$ & 4.4598(9) & 8.081(6) & 13.050(7) & 19.739(14) \\
$\tau$  & 19.38(11) & 64.9(9) & 176(4) & 420(23) \\
\end{tabular}
\end{table}

We determine the dynamic structure factor $C(k,\omega)$ and 
the scaling function $\widetilde{G}(k,t)$ in the 
high-temperature phase $H=0$, $T>T_c$ for small values of $k$---as
we shall see, we are able to reach $k\approx 10/\xi$---by means of 
a large-scale Monte Carlo simulation. We consider the Ising model on 
a cubic lattice, i.e. the Hamiltonian
\begin{equation}
{\cal H} = - \beta \sum_{\langle ij\rangle} \sigma_{i} \sigma_j,
\end{equation}
where $\beta \equiv 1/T$,  $\sigma_i=\pm 1$, and the summation is over 
all nearest-neighbor pairs ${\langle ij\rangle}$. We measure the 
correlation function
\begin{equation}
\widetilde{G}(k,t) = {1\over 3} \sum_{x,y,z} 
     (e^{iqx} + e^{iqy} + e^{iqz}) 
    \langle \sigma_{0,0,0}(t=0)\sigma_{x,y,z}(t)\rangle,
\label{defGtilde-MC}
\end{equation}
for four different values of $L$ and $\beta$: 
(a) $L=64$, $\beta = 0.215$;
(b) $L=64$, $\beta = 0.219$;
(c) $L=128$, $\beta = 0.2204$;
(d) $L=128$, $\beta = 0.221$.
Of course, in Eq.~(\ref{defGtilde-MC}) $q=2\pi n/L$ where $n$ is an integer.
For each $\beta$ and $L$ we first reached equilibrium by running 
20000 Swendsen-Wang iterations, then we collected $N_{\rm it}$ iterations
using the Metropolis algorithm \cite{footnoteMC1}. 
The results of the simulations are reported in Table~\ref{tableMC}. There
we report the number of iterations $N_{\rm it}$, the second-moment 
correlation length $\xi$ (for the $L=128$ lattices we report the more precise 
results of Ref.~\cite{MMPV-02}) and the autocorrelation time $\tau$. 
Note that all lattices have $L/\xi\gtrsim 6$, a condition that 
usually ensures that finite-size effects are reasonably small 
(for static quantities corrections are less than 1\%).

The correlation length $\xi$ has been determined by using a discretized form 
of Eq.~(\ref{xidef}): 
\begin{equation}
\xi^2 = {\chi/F - 1\over 4 \sin^2(\pi/L)},
\end{equation}
where $F = \widetilde{G}(k,0)$ with $k=(2\pi/L,0,0)$. 
The integrated autocorrelation time
$\tau$, and also the autocorrelation times $\tau(k)$ considered below,
have been determined using the self-consistent method of 
Ref.~\cite{MS-86}: 
\begin{equation}
\tau(k) = {1\over2} + \sum_{t=1}^{M(k)} 
   {\widetilde{G}(k,t)\over \widetilde{G}(k,0)},
\end{equation}
where $t$ is the Monte Carlo time in sweeps and the cutoff $M(k)$ is chosen
self-consistently so that $6 \tau(k) < M(k) \le 6 \tau(k) + 1$. 
Since $\widetilde{G}(k,t)$ decays exponentially, this choice makes the 
systematic error due to the truncation small, keeping the statistical variance
small at the same time; see Ref. \cite{MS-86} for a discussion.

First, we check that $\tau\approx \xi^z \approx |T-T_c|^{-z\nu}$.
Using the precise estimate $\beta_c = 0.22165459(10)$ of Ref.~\cite{BST-99},
we obtain from a least-square fit $z= 2.10(2)$ including all data
and $z= 2.11(5)$ discarding the estimate of $\tau$ for lattice (a). This result
is in reasonable agreement with the estimates reported in Sec.~\ref{sec2},
if we take into account that we quote here only the statistical error. 
The systematic error due to corrections to scaling and to neglected finite-size
effects is probably larger.

\begin{figure}
\vspace{0cm}
\centerline{\psfig{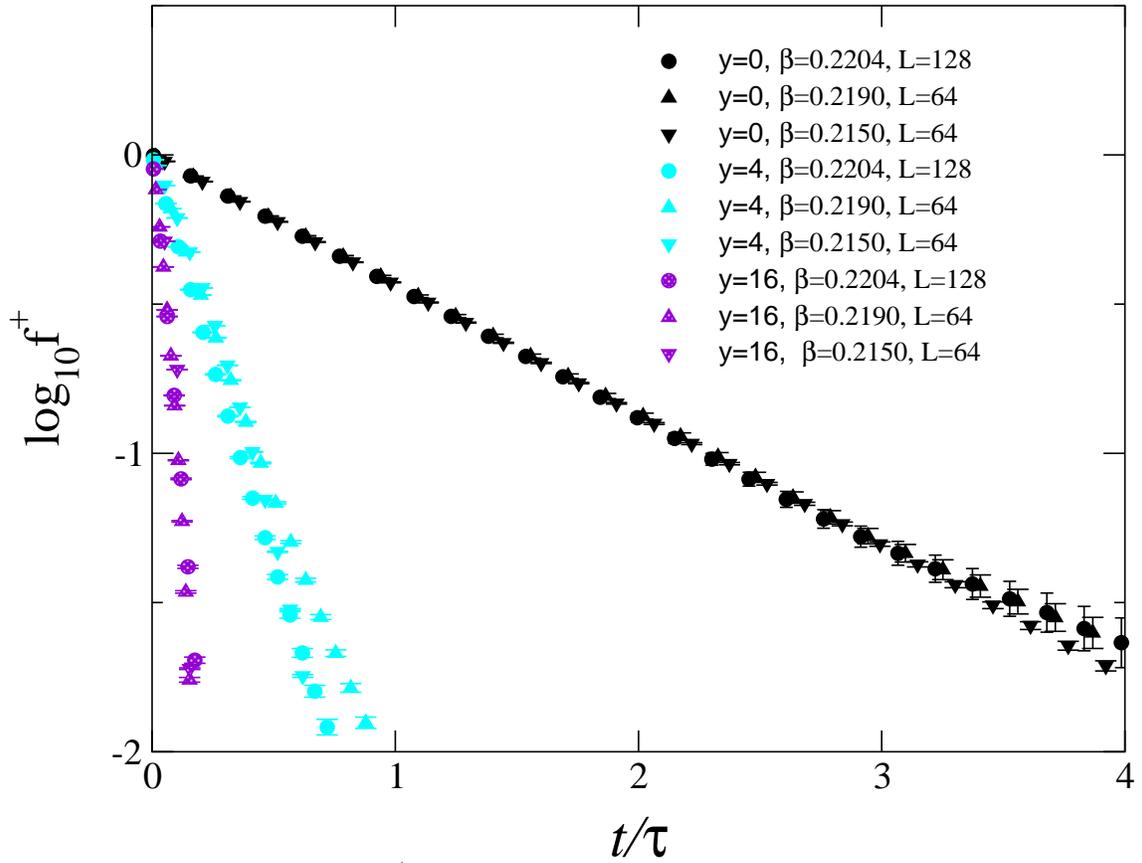}}
\vspace{0cm}
\caption{The scaling function $f^+(y,s)$. We report results for lattices
(a), (b), and (c) and for three different values of $y$.}
\label{figA}
\end{figure} 

Then, we determine the correlation function 
$\widetilde{G}(k,t)$. In Fig.~\ref{figA} we report the function
\begin{equation}
f^+(y,s) \equiv  {g^+(y,0)\over g^+(y,s)} \approx 
   {\widetilde{G}(k,t)\over \widetilde{G}(k,0)},
\end{equation}
for three different values of $y\equiv k^2 \xi^2$, $y=0,4,16$, as computed 
from lattices (a), (b), and (c). We have not included the results 
for lattice (d), because they have much larger errors. In order
to obtain $\widetilde{G}(k,t)$ for a given $k\not=2 \pi n/L$, we have performed
a linear interpolation, using two nearby values of $k$. First, we observe
reasonable scaling: corrections due to the finite values of $\xi$ and $L$
are under control, although they increase as $y$ increases. 
For $y=0$ the results for the three different lattices agree within a few 
percent, while for larger values of $y$ we observe larger discrepancies.
In particular, for $y=4$ and $y=16$, the estimates of $f^+(y,s)$ obtained 
from lattice (b) are always larger than those obtained from (a) and (c),
the discrepancy being of order 20\% when $f^+(y,s)\approx 10^{-1}$ and 
80\% when $f^+(y,s)\approx 10^{-2}$. These differences are probably finite-size
effects, since (a) and (c) have $L/\xi \gtrsim 10$, while $L/\xi \approx 8$
for (b). 

It is also remarkable that the plot of $\ln f^+(y,s)$  is 
a straight line,
indicating that $f^+(y,s)$ is quite precisely a pure exponential. 
No deviations can be observed in Fig.~\ref{figA}. Therefore, 
\begin{equation}
  \widetilde{G}(k,t) \approx \widetilde{G}(k,0) 
   \exp\left[-t/\tau_{\rm exp}(k)\right],
\label{Gtilde-Gauss}
\end{equation}
within the precision of our results. 
This behavior appears to be well satisfied in the 
region that we can safely investigate, i.e. 
$1/10 \lesssim t/\tau(k) \lesssim 4$  and
$k \xi \lesssim 5$. Therefore, the dynamic structure factor is 
well approximated by a Lorentzian in the region of not too large frequencies, 
i.e. for $\omega \tau(k) \ltapprox 10$. 

\begin{figure}
\vspace{0cm}
\centerline{\psfig{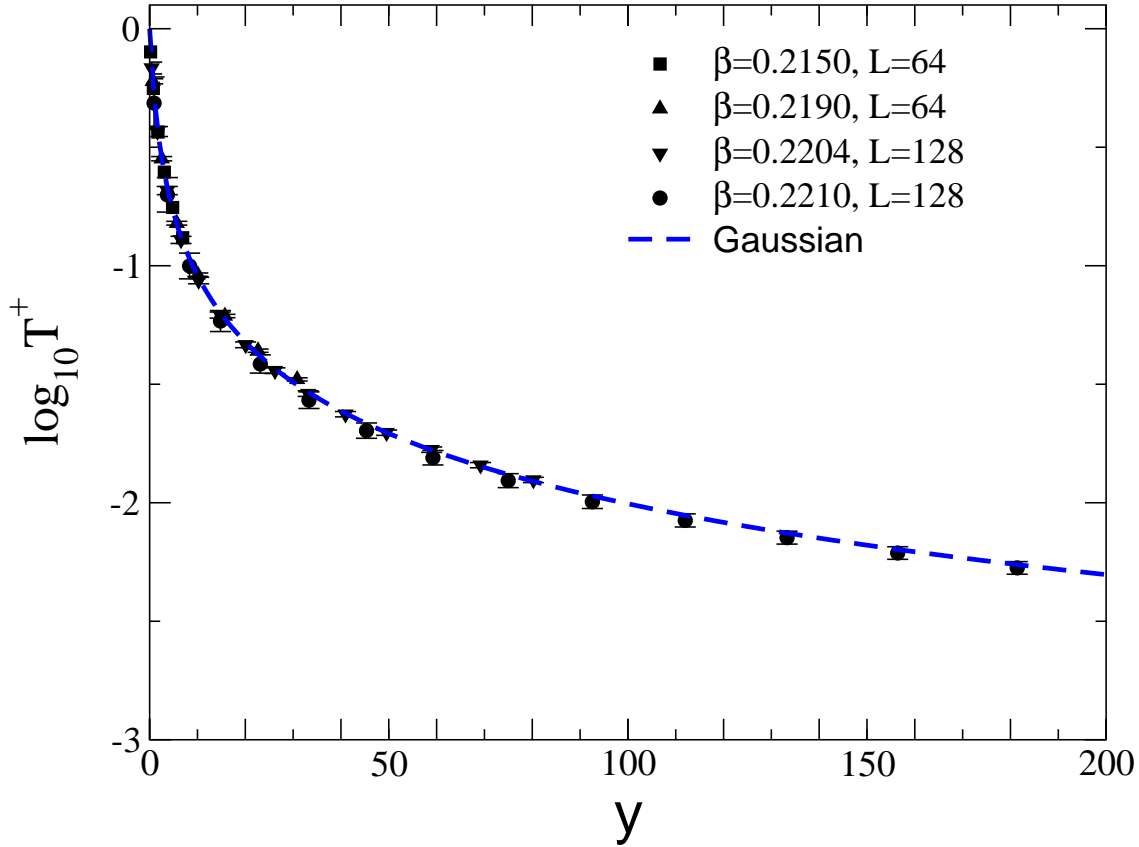}}
\vspace{0cm}
\caption{Scaling plot of ${\cal T}(y)$ vs $y\equiv k^2 \xi^2$.}
\label{figT}
\end{figure}

Then, we consider the scaling function ${\cal T}(y)$ that encodes the 
$k$-dependence of $\tau(k)$. In Fig.~\ref{figT} we report our numerical 
results. Again, we observe good scaling up to quite large values of 
$y$. In the figure, we also report the Gaussian prediction
${\cal T}(y) = 1/(1 + y)$. It can be seen that the Gaussian approximation
describes very well the numerical data. This result should have been 
expected on the basis of the results of Sec. \ref{sec3} where we showed 
that the deviations from a Gaussian behavior are very small in the 
small-$y$ regime $y\ltapprox 9$, and should remain small even for larger $y$.
For instance, using the data with largest $y$ reported in Fig.~\ref{figT},
we estimate
${\cal T}(y) = 0.0053(3)$ for $y=181$, to be compared with the 
Gaussian prediction $0.0055$. Thus, in the range $y\lesssim 200$ the 
discrepancy should be at most 4-10\%.

\begin{figure}
\vspace{0cm}
\centerline{\psfig{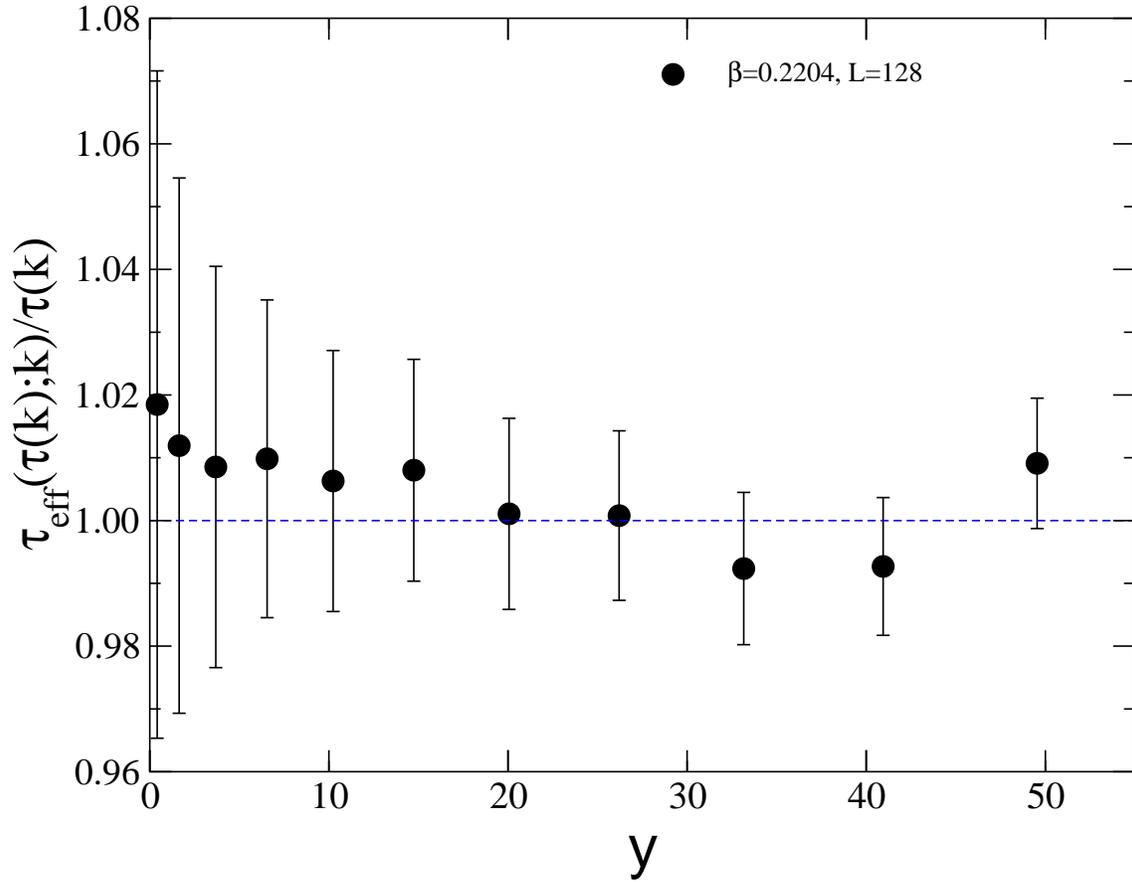}}
\vspace{0cm}
\caption{Ratio $\tau_{\rm eff}(t;k)/\tau(k)$ vs $y\equiv k^2 \xi^2$,
for $t = \tau(k)$. Results for lattice (c), $L=128$, $\beta = 0.2204$.}
\label{figTexp}
\end{figure}

Finally, we consider the function ${\cal T}^+_{\rm exp}(y)$. In order to 
compute $\tau_{\rm exp}(k)$ we define an effective quantity
\begin{equation}
\tau_{\rm eff}(t;k) \equiv 
  - \left[ \ln{\widetilde{G}(k,t+1)\over \widetilde{G}(k,t)}\right]^{-1}.
\end{equation}
The exponential autocorrelation time $\tau_{\rm exp}(k)$ is 
obtained from $\tau_{\rm eff}(t;k)$ by letting $t$ go to infinity. 
In practice, we can only compute $\tau_{\rm eff}(t;k)$ up to 
$t$ of the order of (1-2)$\times\tau(k)$ since errors increase rapidly. 
In Fig.~\ref{figTexp} we report the ratio $\tau_{\rm eff}(t;k)/\tau(k)$ 
for $t = \tau(k)$ for lattice (c) which is the only one that allows us to 
reach large values of $y$. We observe $\tau_{\rm eff}(t;k)\approx \tau(k)$
within the precision of our results. It is tempting to conclude that 
$\tau_{\rm exp}(k) \approx \tau(k)$ for $y < 50$, but this is in contrast
with the theoretical results of Sec. \ref{sec3.B}. Indeed, we showed there 
that $\tau_{\rm exp}(k) \approx \tau(k)$ with very small corrections for 
$y<3$, but we noticed that this relation breaks down for larger values of 
$y$. For instance, for $y>3$, our two-loop calculation gives 
$\tau_{\rm exp}(k)/\tau(k) = (3 + 3y)/(9 + y)$ which is significantly larger 
than 1 
for $y> 3$. As we already discussed this prediction should not be taken 
seriously, unless $y$ is close to 3, since other singularities should be 
present,
and indeed we expect $\tau_{\rm exp}(k)/\tau(k)$ to diverge as $k\to \infty$.
Therefore, our numerical data show that the asymptotic large-$t$ behavior 
sets in only for large values of $t$, i.e. for $t\gg \tau(k)$, where 
the correlation function $\widetilde{G}(k,t)$ is very small. 
Therefore, even if Eq.~(\ref{Gtilde-Gauss}) breaks down for 
$y\gtrsim 3$ and $t$ large, it still represents a very good 
approximation (even for $y\approx 50$) for the values of $t$ for 
which $\widetilde{G}(k,t)$ is sizeable.

\section*{Acknowledgements}

V.M.-M. is a {\em Ramon y Cajal} research fellow and is partly supported
by MCyT (Spain), project Nos. FPA2001-1813 and FPA2000-0956.

\appendix

\section{Integrals entering the field-theoretical calculations}
\label{appa}

In this appendix we report some integrals that enter
the perturbative field-theoretical calculations.

In the two-loop calculation of the response scaling function in 
the high-temperature phase,
cf. Eq.~(\ref{ayw}), one needs to compute the function
\begin{equation}
A(k^2,\omega) =
{2\over 27} N_d^{-2} \left[ I(k,\omega) - I(0,0) \right]  ,
\end{equation}
with~(dimensional regularization near four dimensions is understood)
\begin{equation}
I(k,\omega) =  \int^{\infty}_{0} dt
e^{ i \omega t}
\int \frac{d^dp_1}{(2\pi)^d}\frac{d^dp_2}{(2\pi)^d}
\prod_{i=1}^3\frac{1}{p_i^2+1} e^{-t \sum_i (p_i^2 + 1) } ,
\label{eqA2}
\end{equation}
where $p_3 = k - p_1 - p_2$, and $N_d = 2/[(4\pi)^{d/2} \Gamma(d/2)]$.
The integral $I(k,\omega)$ can be written in the form
\begin{equation}
\label{intcom}
N_d^{-2} I(k,\omega)=
\frac{1}{3} {1\over (4 \pi)^d N_d^2}
\int_0^1 t^2 dt \int_0^{\infty} ds e^{-s}
s^{3-d} e^{i \omega s \frac{1-t}{3}} 
\int_0^1 u du dv \frac{e^{-s\frac{Q}{\Delta}k^2}}
{\Delta^{d/2}}, 
\end{equation}
where 
\begin{eqnarray}
&&\Delta= t^2u [1-u+uv(1-v)]+\frac{1}{3}(1-t^2),\\
&&Q=t^3u^2(1-u)v(1-v)+\frac{1-t}{3}t^2u[1-u+uv(1-v)]+\frac{(1-t)^2}{9} t
+\frac{(1-t)^3}{27} .
\end{eqnarray}
We will also need the singularity structure of the $A(k^2,\omega)$. 
For this purpose, we will determine the large-$t$ behavior of 
$\tilde{A}(k^2,t)$ which is the Fourier transform with respect to 
$\omega$ of $A(k^2,\omega)$. This behavior can easily be derived from 
Eq.~(\ref{eqA2}). Setting $p_1 = k/3 + q_1/\sqrt{t}$, 
$p_2 = k/3 + q_2/\sqrt{t}$, $p_3 = k/3 + q_3/\sqrt{t}$, we obtain that for 
$t\to\infty$, 
\begin{eqnarray}
\tilde{A}(k^2,t)&\approx &\frac{2}{27} N_d^{-2} t^{-d}e^{-k^2t/3-3t}
\frac{1}{(k^2/9+1)^3}\int \frac{d^dq_1}{(2\pi)^d}\frac{d^dq_2}{(2\pi)^d}
e^{-(q_1^2+q_2^2+q_3^2)}[1+O(t^{-1/2})]
\nonumber\\
&= &\frac{1}{(k^2/9+1)^3}\frac{(1/3)^{d/2}}{54}
\Gamma^2(d/2)t^{-d}e^{-k^2t/3-3t}[1+O(t^{-1/2})]
\nonumber\\
&\approx &\frac{1}{(k^2+9)^3} \frac{3e^{-k^2t/3-3t}}{2 t^4}[1+O(t^{-1/2})]
+O(\epsilon)
\end{eqnarray}
This result implies the presence of a branching cut in $A(k^2,\omega)$
starting at $\omega = - i(3 + k^2/3)$.

The one-loop expression of the response function in the ordered 
phase, cf. Eq.~(\ref{byw}), is written in terms of the function
\begin{equation}
B(k^2,\omega) =  N_d^{-1}\left[ J(k,\omega) - J(0,0)\right],
\end{equation}
where
\begin{equation}
J(k,\omega) =  \int^{\infty}_{0} dt
e^{ i \omega t}
\int \frac{d^dp_1}{(2\pi)^d}
\prod_{i=1}^2\frac{1}{p_i^2+1} e^{-t \sum_i^2 (p_i^2 + 1) } ,
\end{equation}
with $p_2 = k - p_1$.
The function $B(k^2,\omega)$ can be written in the form
\begin{equation}
B(k^2,\omega)={1\over 2} \int_0^1 t\, dt \,du
{i 2 \omega (1-t) - [ 1 - t^2 + 4 t^2 u (1-u)] k^2  
\over 4 + [ 1 - t^2 + 4 t^2 u (1-u)] k^2 -
2 i \omega (1-t)}.
\end{equation}
Such an integral can be computed exactly obtaining
\begin{eqnarray}
B(k^2,\omega)&=&
 -{1\over2} +\frac{1}{k^2}
   \ln \frac{4 + 2 i \omega + k^2}{4 - 2 i \omega + k^2}+
\frac{2 i}{\omega}\,\sqrt{\frac{4+k^2}{k^2}}
 \ln \frac{\sqrt{4+k^2}-\sqrt{k^2}}{\sqrt{4+k^2}+ \sqrt{k^2}}
\nonumber\\
&&  -\frac{1}{k^2} \ln \frac{4\,\omega^2 +(4+k^2)^2}{16} +
     \frac{2}{\omega k^2} F \left(\ln \frac{F-\omega}{F+\omega}+
     \ln \frac{F+ \omega + i k^2 }{F-\omega - i k^2} \right)
\end{eqnarray}
with
$ F\equiv\sqrt{\omega^2 + 2 i \omega k^2 - k^2 (4 + k^2)} $.
It is easy to see using this exact expression or repeating the argument 
presented for $A(k^2,\omega)$ that $B(k^2,\omega)$ is singular for 
$\omega = - i(2 + k^2/2)$.



\begin{thebibliography}{199}

\bibitem[*]{PC-email}
Email address: Pasquale.Calabrese@df.unipi.it

\bibitem[\dag]{VM-email}
Email address: Victor@lattice.fis.ucm.es

\bibitem[\ddag]{AP-email}
Email address: Andrea.Pelissetto@roma1.infn.it

\bibitem[\S]{EV-email}
Email address: Ettore.Vicari@df.unipi.it

\bibitem{HH-77}
P. C. Hohenberg and B. I. Halperin,
Rev. Mod. Phys. {\bf 49}, 435 (1977).

\bibitem{HJSSV-98}
M. A. Halasz, A. D. Jackson, R. E. Shrock, M. A. Stephanov,
and J. J. Verbaarschot, Phys. Rev. D {\bf 58}, 096007 (1998).

\bibitem{BR-99}
J. Berges and K. Rajagopal, Nucl. Phys. B {\bf 538}, 215 (1999).

\bibitem{BR-00}
B. Berdnikov and K. Rajagopal,
Phys. Rev. D {\bf 61}, 105017 (2000).

\bibitem{BPSS-01}
Sz. Bors\'anyi, A. Patk\'os, D. Sexty, and Zs. Sz\'ep,
Phys. Rev. D {\bf 64}, 125011 (2001).

\bibitem{CG-02}
P. Calabrese and A. Gambassi, 
Phys. Rev. E {\bf 65}, 066120 (2002); 
E {\bf 66}, 066101 (2002).

\bibitem{MSR-73}
P. C. Martin, E. D. Siggia, and H. A. Rose,
Phys. Rev. A {\bf 8}, 423 (1973).

\bibitem{BJW-76}
R. Bausch, H. K. Janssen and H. Wagner,
Z. Physik B {\bf 24}, 113 (1976).

\bibitem{DP-78}
C. De Dominicis and L. Peliti,
Phys. Rev. B {\bf 18}, 353 (1978).

\bibitem{GZ-98} R.~Guida and J.~Zinn-Justin,
J.~Phys. A  {\bf 31}  (1998) 8103.

\bibitem{CPRV-99}
M.~Campostrini, A.~Pelissetto, P.~Rossi, and E.~Vicari,
Phys. Rev. E {\bf 60}, 3526 (1999).

\bibitem{BST-99}
H.~W.~J.~Bl\"ote, L. N. Shchur, and A. L. Talapov,
Int. J. Mod. Phys. C  {\bf 10}, 1137 (1999).

\bibitem{Hasenbusch-01}
M.~Hasenbusch, Int. J. Mod. Phys. C  {\bf 12}, 911 (2001).

\bibitem{BC-02}
P.~Butera and M.~Comi,
Phys. Rev. B  {\bf 65}, 144431 (2002).

\bibitem{CPRV-02}
M. Campostrini, A. Pelissetto, P. Rossi, and E. Vicari,
Phys. Rev. E {\bf 65}, 066127 (2002).

\bibitem{review}
A. Pelissetto and E. Vicari,
Phys. Rep. {\bf 368}, 549 (2002).

\bibitem{BDJZ-81}
R. Bausch, V. Dohm, H. K. Janssen, and R. P. K. Zia,
Phys. Rev. Lett. {\bf 47}, 1837 (1981).

\bibitem{AV-84}
N. V. Antonov and A. N. Vasil'ev,
Teor. Mat. Fiz. {\bf 60}, 59 (1984)
[Theor. Math. Phys. {\bf 60}, 671 (1984)].

\bibitem{PIF-97}
V. V. Prudnikov, A. V. Ivanov, and A. A. Fedorenko,
Pis'ma Zh. Eksp. Teor. Fiz. {\bf 66}, 793 (1997)
[JETP Lett. {\bf 66}, 835 (1997)].

\bibitem{WL-91}
S. Wansleben and D. P. Landau,
Phys. Rev. B {\bf 43}, 6006 (1991).

\bibitem{Heuer-92}
H. Heuer, J. Phys. A {\bf 25}, L567 (1992).


\bibitem{Gro-95}
U. Gropengiesser, Physica A {\bf 213}, 308 (1995).

\bibitem{Grassberger-95}
P. Grassberger, Physica A {\bf 214}, 547 (1995);
(E) A {\bf 217}, 227 (1995).

\bibitem{SK-96}
D. Stauffer and R. Knecht,
Int. J. Mod. Phys. C {\bf 7}, 893 (1996).

\bibitem{Stauffer-97}
D. Stauffer,
Physica A {\bf 244}, 344 (1997).

\bibitem{ZZ-98}
G. P. Zheng and J. X. Zhang,
Phys. Rev. E {\bf 58}, R1187 (1998).

\bibitem{JMSZ-99}
A. Jaster, J. Mainville, L. Sch\"ulke, and B. Zheng,
J. Phys. A {\bf 32}, 1395 (1999).

\bibitem{Belanger-etal-88}
D. P. Belanger, B. Farago, V. Jaccarino, A. R. King,
C. Lartigue, F. Mezei,
J. Phys. (France) Colloque C8, {\bf 49}, 1229 (1988).

\bibitem{Hutchings-etal-72}
M. T. Hutchings, M. P. Schulhof, and H. J. Guggenheim,
Phys. Rev. B {\bf 5}, 154 (1972).

\bibitem{FB-67}
M.~E.~Fisher and R.~J.~Burford,
Phys.\ Rev.\  {\bf 156}, 583 (1967).

\bibitem{TF-75}
H.~B.~Tarko and M.~E.~Fisher,
Phys.\ Rev.\ Lett.\  {\bf 31}, 926 (1973);
Phys.\ Rev.\ B  {\bf 11}, 1217 (1975).

\bibitem{CPRV-98} 
M.~Campostrini, A.~Pelissetto, P.~Rossi, and E.~Vicari,
Phys.\ Rev.\ E {\bf 57}, 184 (1998).

\bibitem{MMPV-02}
V.~Mart\'{\i}n-Mayor, A. Pelissetto, and E. Vicari,
Phys. Rev. E {\bf 66}, 026112 (2002).

\bibitem{CDK-75}
M. Combescot, M. Droz, and J. M. Kosterlitz,
Phys. Rev. B {\bf 11}, 4661 (1975).

\bibitem{Bray-76}
A. J. Bray, Phys. Rev. B {\bf 14}, 1248 (1976).

\bibitem{GZ-97}
R.~Guida and J.~Zinn-Justin,
Nucl.\ Phys.\ B  {\bf 489}, 626 (1997).

\bibitem{EFS-02}
J. Engels, L. Fromme, and M. Seniuch, 
``Numerical equation of state from an improved three-dimensional Ising model,"
e-print cond-mat/0209492.

\bibitem{intom}
Here we use the fact that the integral
$\int \case{d\omega}{2\pi} |\omega|^x {\rm exp}(i\omega t)$
is given by $-\pi^{-1} \sin (\pi x /2) \Gamma(1+x) |t|^{-1-x}$ for $x > - 1$.

\bibitem{FS-75}
R. A. Ferrell and D. J. Scalapino,
Phys. Rev. Lett.  {\bf 34}, 200 (1975).

\bibitem{foot-3pc}
Note that, beside the three-particle cut, $\widetilde{G}(k,0)$ shows an 
additional singularity, the one-particle pole at $k=\pm i m_{\rm exp}$. 
However, such a point is a simple zero---therefore {\em not} a 
singularity---of the inverse $[\widetilde{G}(k,0)]^{-1}$. 

\bibitem{FZ-98}
M.~E.~Fisher and S.-Y.~Zinn, J.\ Phys. A {\bf 31}, L629 (1998).


\bibitem{footnoteMC1} We used the Metropolis algorithm with {\em sequential }
updating of the sites. At variance with the Metropolis algorithm with 
random updating, this dynamics is not reversible (it does not satisfy detailed 
balance). Nonetheless, it is commonly 
accepted that in the critical limit these two dynamics should both belong to 
the dynamic universality class of model A that describes a 
{\em reversible} diffusive dynamics without conservation laws. 

\bibitem{MS-86}
N. Madras and A. D. Sokal, J. Stat. Phys.  {\bf 50}, 109 (1988).
\end{thebibliography}
\end{document}